\documentclass[preprint,endfloats,prl,aps,showpacs]{revtex4}

\usepackage{graphicx}
\usepackage{amsmath}
\usepackage{amssymb}
\usepackage{multirow}

\begin{document}

\title{Energy-dependent tunnelling from few-electron dynamic quantum dots.}

\author{M.~R.~Astley,$^{1,2}$ M.~Kataoka,$^{1}$ C.~J.~B.~Ford,$^{1}$ C.~H.~W.~Barnes,$^{1}$ D.~Anderson,$^{1}$ G.~A.~C.~Jones,$^{1}$ I.~Farrer,$^{1}$ D.~A.~Ritchie,$^{1}$ and M.~Pepper.$^{1,2}$}

\affiliation{$^{1}$Cavendish Laboratory, University of Cambridge, J.~J.~Thomson Avenue, Cambridge CB3 0HE, United Kingdom\\ $^{2}$Toshiba Research
Europe Limited, Cambridge Research Laboratory, 260 Cambridge Science Park, Cambridge CB4 0WE, United Kingdom}


\begin{abstract}
We measure the electron escape-rate from surface-acoustic-wave dynamic quantum dots (QDs) through a tunnel barrier.  Rate-equations are used to extract the tunnelling rates, which change by an order of magnitude with tunnel-barrier gate voltage. We find that the tunnelling rates depend on the number of electrons in each dynamic QD because of Coulomb energy. By comparing this dependence to a saddle-point-potential model, the addition energies of the second and third electron in each dynamic QD are estimated. The scale ($\sim$ a few meV) is comparable to those in static QDs as expected.

\end{abstract}

\maketitle

Quantum dots (QDs) in semiconductor systems, where electrons are confined into zero-dimensional states, have been the object of much recent
attention \cite{Kouwenhoven1998, Kouwenhoven2001}. In a gate-defined quantum dot the number of electrons can be reduced down to one
\cite{Tarucha1996, Ciorga2000}; such single-electron QDs may form the basis of qubits in quantum computation schemes \cite{Loss1998, Vandersypen}.
High frequency operations on QD systems have been used to observe fundamental electronic phenomena such as coherent charge oscillations
\cite{Hayashi2003}, single- and multiple-spin dynamics \cite{Elzerman2004, Petta2005, Koppens2006}, excited state spectra \cite{Elzerman2004b},
and elastic tunnelling behaviour \cite{Maclean2007}, and will be necessary for quantum computation applications in semiconductor systems.

In typical QD experiments, the QDs were defined by \emph{static} surface gates, and high frequency operations were achieved by applying voltage
pulses to the gates. However, an alternative method has received recent attention: to use a \emph{dynamic} QD defined by a surface acoustic wave
(SAW) where high frequency operations are performed by moving the QD past static surface gates at a high velocity \cite{Barnes2000,
Rodriquez2005}. Because GaAs is piezo-electric, the strain wave of a SAW on a GaAs/AlGaAs heterostructure is accompanied by an electric potential
modulation, which forms a series of one- or few-electron dynamic QDs in an empty quasi-one-dimensional channel \cite{Shilton1996,
Talyanskii1997}. Previous experiments have attempted to observe interactions in dynamic QDs defined by SAWs \cite{Kataoka2006}, but, to our
knowledge, the tunnelling behaviour necessary to observe complex quantum phenomena has not been seen.

In this letter we report measurements of the non-equilibrium escape rate from one- and few-electron dynamic QDs defined by a SAW. This measurement has been carried out in static quantum dots over second \cite{Cooper2000} and millisecond \cite{Maclean2007} timescales, but the dynamic QD arrangement allows us to directly observe electron tunnelling on sub-nanosecond timescales. The SAW-defined dynamic QDs
carry electrons along the channel to a tunnel barrier, where the electrons can escape from the QD into a neighbouring two-dimensional electron gas
(2DEG).
Observation of the tunnelling current allows us to determine the tunnel rate out of the dynamic QD, which is found to depend on the number
of electrons in the dot. By fitting these rates to a simple model, we determine the addition energy of the dynamic QD. This is, to our knowledge, the first direct measurement of dynamic QD energies.

\begin{figure}
\includegraphics[width=0.5\textwidth]{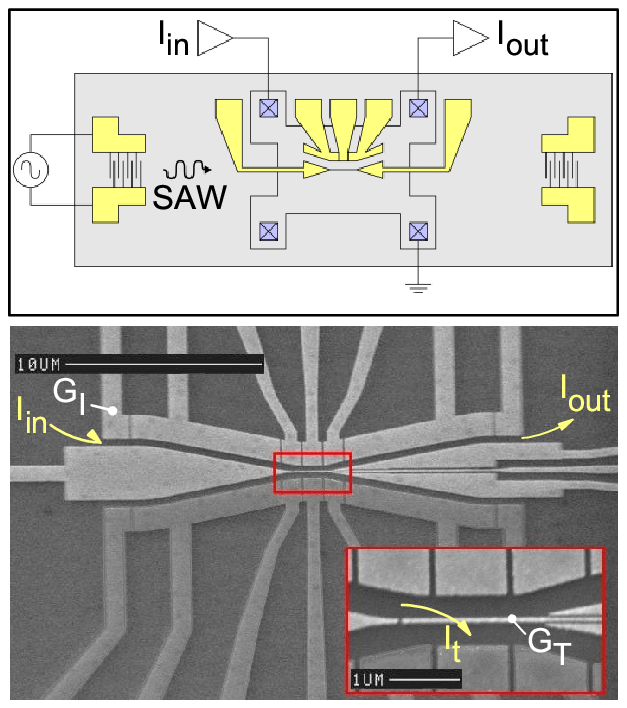}
\caption{(colour online). Upper panel: Schematic of the device design. Lower panel: Electron micrograph of the device's surface gates. The dark shaded gates were grounded.} \label{device}
\end{figure}
The device was made using a modulation-doped GaAs/AlGaAs heterostructure, which had a 2DEG 97~nm below the surface with a mobility of
160~$\mathrm{m^2/Vs}$ and a carrier density of 1.8~$\mathrm{\times10^{15}m^{-2}}$ in the dark. NiCr/Au surface gates (shown in Fig.~\ref{device})
deplete the 2DEG under negative bias, to create the SAW channel and tunnelling region.
The SAW was generated by applying a 11.1~dBm microwave signal from an Agilent 8648D signal generator to a transducer, made of 70 pairs of interdigitated
fingers with a period of 1~$\mathrm{\mu m}$, situated 2.5~mm from the device. Based on measurements from a device that underwent identical processing \cite{Schneble2006}, we predict that this would result in a SAW amplitude of $\sim 50\,\mathrm{meV}$ peak-to-peak.
The microwave power was pulse-modulated using a Tektronix PG5110
pulse generator, with a duty ratio of $10\,\mathrm{\mu s}\,:\,500\,\mathrm{\mu s}$ to minimise sample heating \cite{Schneble2006}. Measurements were
carried out in a $\mathrm{^3He}$ cryostat with a base temperature of 270~mK.

The injector gate ($\mathrm{G_I}$) is used to control the number of electrons that can enter the SAW channel. At sufficient SAW power the injected
current becomes quantised to $I_\mathrm{in} = Nef$, where $e$ is the electron charge and $f$ is the SAW frequency. In this regime each SAW minimum
forms a dynamic QD that contains $N$ electrons, moving through the channel at the SAW velocity ($\sim2800\,\mathrm{ms^{-1}}$). When the dot is
alongside the tunnel barrier gate ($\mathrm{G_T}$), the electrons are coupled to the reservoir and tunnel out of the dot; this tunnelling process
is described by $\Gamma_n$, the rate at which an electron leaves an $n$-electron dynamic QD (Note that we use $N$ for the number of initially injected electrons in each QD, whereas $n$ is the number of electrons in a QD in the tunnel barrier region). Escape of electrons from the dot means the current
$I_\mathrm{out}$ coming out of the channel is reduced by a tunnelling current $I_\mathrm{t}$. The effective length of the tunnel barrier
can be estimated by solving Laplace's equation for the device's surface gate voltages \cite{Davies1995} as $\sim 1.6\,\mathrm{\mu m}$, meaning the dynamic QD is coupled to the reservoir for a tunnelling duration ($\tau$) of about $600\,\mathrm{ps}$. Although the exact barrier length may be smaller than this because of impurity or disorder potentials, we determine $\Gamma \tau$ in our analysis, so uncertainty in the exact value of $\tau$ does not effect our results. The remaining gates that define the channel are held at constant voltage throughout the experiment; these voltages have been carefully tuned to minimise any  potential gradients in the channel, as large potential gradients could cause a loss of confinement in the dynamic QDs and lead to fluctuations from the initialised electron number $N$.

\begin{figure}
\includegraphics[width=0.5\textwidth]{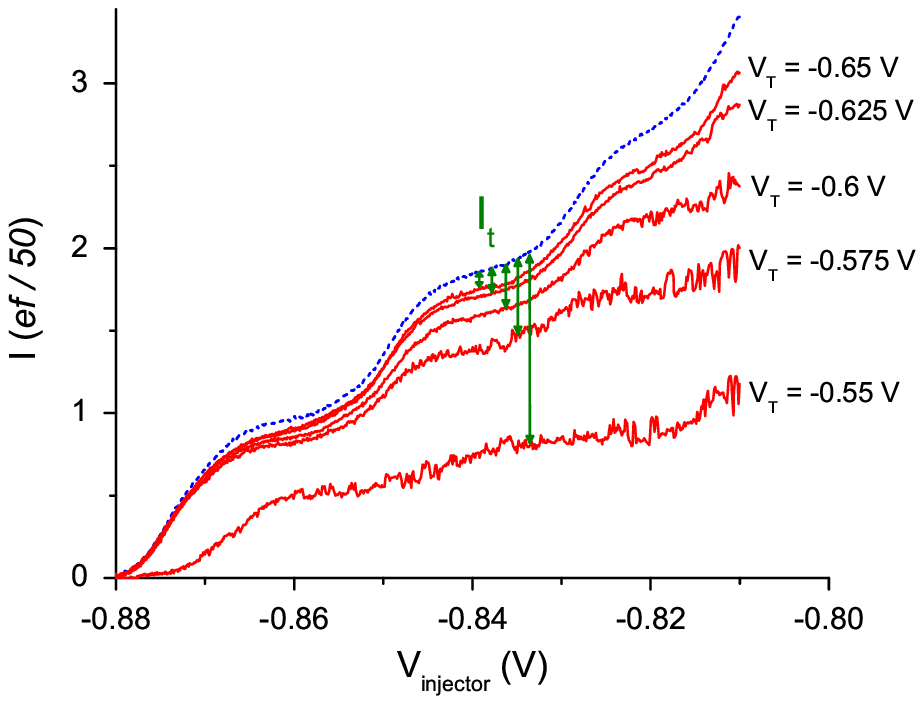}
\caption{(colour online). $I_\mathrm{in}$ (dotted line) and $I_\mathrm{out}$ (solid lines) dependence on injector gate voltage, for a range of
barrier gate voltages. Plateaux occur when an integer number ($N$) of electrons occur in each SAW minimum. $I_\mathrm{t}$ is the difference between
the two curves.} \label{Idp}
\end{figure}

The dotted line in Fig.~\ref{Idp} shows $I_\mathrm{in}$ as a function of the voltage applied to the injector gate. The first three quantised
plateaux can be seen at multiples of $8.7\,\mathrm{pA}$, which is $Nef$ reduced by the $1\,:\,50$ pulse ratio used \cite{Kataokab}. The solid lines
show $I_\mathrm{out}$ for a range of voltages ($V_\mathrm{T}$) applied to the tunnel barrier---the less negative the barrier voltage, the higher
the rate of tunnelling out of the channel, thus the lower the value of $I_\mathrm{out}$. The tunnelling current $I_\mathrm{t}$ is deduced from the
difference between $I_\mathrm{in}$ and $I_\mathrm{out}$.

\begin{figure}
\includegraphics[width=0.5\textwidth]{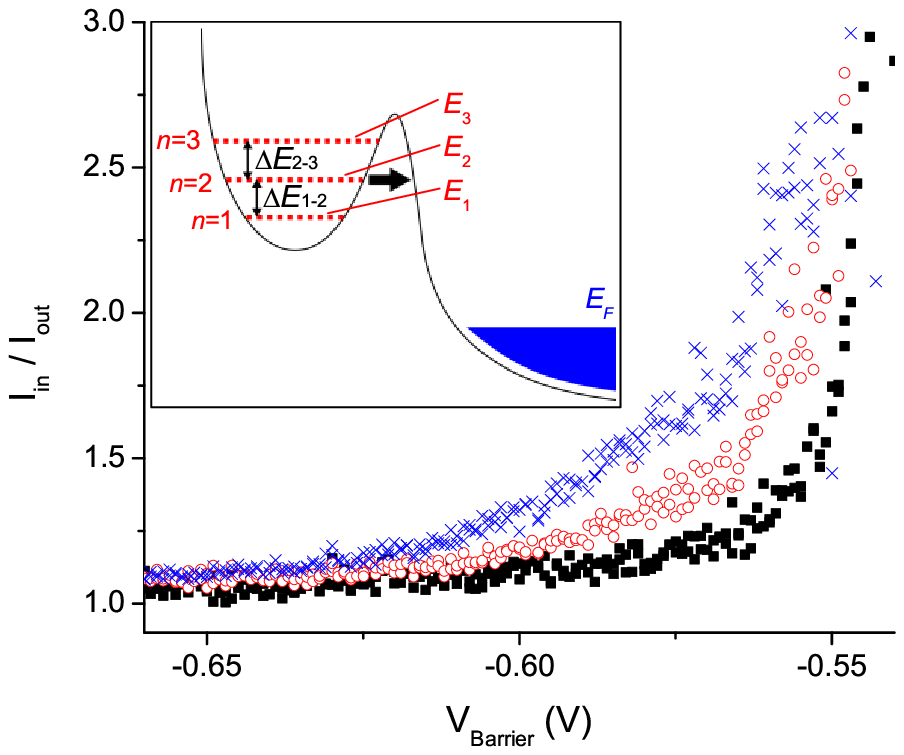}
\caption{(colour online). The ratios $I_\mathrm{in}/I_\mathrm{out}$ as a function of barrier gate voltage, taken from the $I_\mathrm{in}$ plateau
corresponding to $N=1$ ($\blacksquare$), $N=2$ ($\bigcirc$) and $N=3$ ($\times$). Inset: Schematic of tunnelling of electrons from the dynamic QD across the barrier into the reservoir---energies of the electrons within the dot are dependent on $n$, leading to $n$-dependent tunnelling rates.}
\label{ratios}
\end{figure}

In previous SAW measurements, it was not possible to demonstrate that electrons were confined in a dynamic QD for the entire length of a long
SAW channel (an essential feature of proposed SAW quantum circuits). An alternative possibility was that quantised charge pumping occurred at a
microconstriction, but subsequently electrons could escape from the dot and freely move along the channel. In our device, if electrons were not confined in dynamic QDs but were free to move in an open channel, we would expect that adding up to three electrons in a SAW cycle would not change the energy of the system. Hence such behaviour would be unobservable and the ratio $I_\mathrm{in}/I_\mathrm{out}$ would be independent of $N$. On the other hand, if electron confinement is maintained, the energy state of the dot can vary by several meV depending on the number of electrons present and the size of the confinement potential, and thus the tunnelling rate and therefore $I_\mathrm{in}/I_\mathrm{out}$ should be number dependent. In
Fig.~\ref{ratios} $I_\mathrm{in}/I_\mathrm{out}$ is shown as a function of barrier-gate voltage for $N=1,\,2,\,3$.
The ratio $I_\mathrm{in}/I_\mathrm{out}$ is strongly dependent on $N$, indicating that the dynamic QD model correctly describes our system for at least the whole tunnel barrier region.

\begin{figure}
\includegraphics[width=0.5\textwidth]{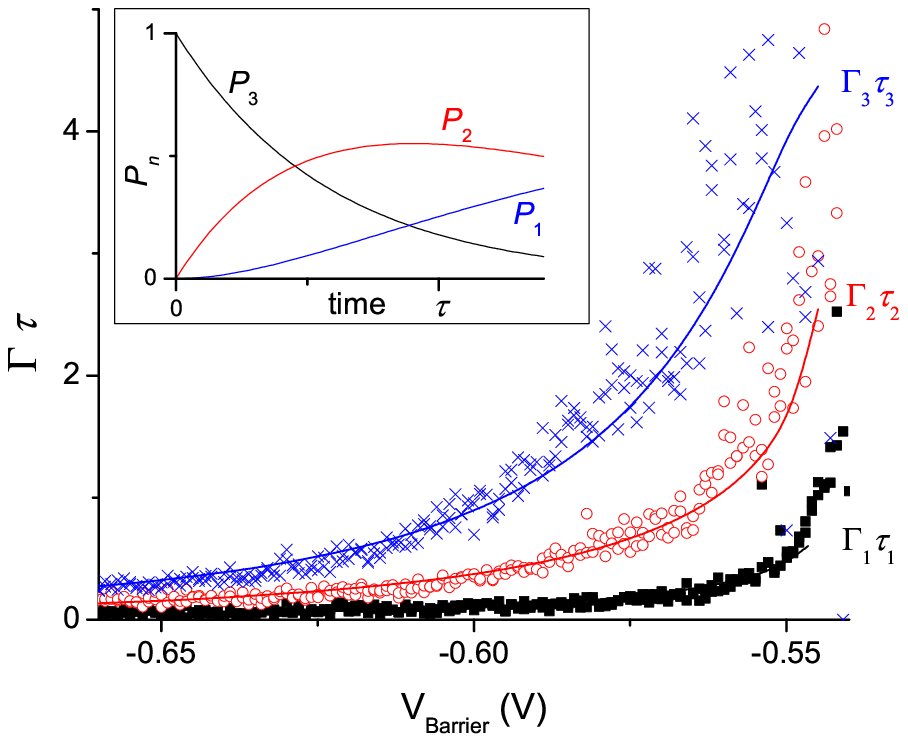}
\caption{(colour online). Dependence of the calculated tunnelling rates $\Gamma_n$ on barrier gate voltage for one ($\blacksquare$), two ($\bigcirc$) and three
($\times$) electrons in the dynamic QD, normalised by the tunnelling time $\tau_n$. The solid lines show fits based on the tunnelling probability of
non-interacting electrons incident on a saddle-point potential, as described in the text. Inset: Example of the time-evolution of $P_n$ for $3ef$ injection, using
tunnel rates for $V_\mathrm{T}=-0.575$~V.} \label{Gtdp}
\end{figure}

Control of the tunnelling rate of electrons leaving a QD is needed for understanding and manipulating the quantum states within the dot. We can
deduce the tunnelling rate $\Gamma_n$ of an $n$-electron dynamic quantum dot by comparing our measurements with rate equations. Within the
tunnelling region, the probability ($P_n$) for having $n$ electrons in the dot varies with time according to $\frac{dP_n}{dt}=\Gamma_{n+1} P_{n+1}
- \Gamma_n P_n$ (we treat each dynamic QD as undergoing an independent tunnelling event---there is a $\sim1\,\mathrm{\mu m}$ $\sim50$~meV barrier between electrons in neighbouring dots so there will be no wavefunction overlap, and the Coulomb energy of two electrons $\sim1\,\mathrm{\mu m}$ apart is only $\sim100\,\mathrm{\mu eV}$ which should have little effect). Assuming that the tunnel rates $\Gamma_n$ remain constant over the duration of tunnelling $\tau_n$, that on the
$I_\mathrm{in}=Nef$ plateau there are exactly $N$ electrons in each SAW minimum, and that no electrons are able to tunnel back into the dot, $I_\mathrm{out}=ef\sum_{n=1}^N nP_n$ can be calculated as:
\begin{align*}
I_\mathrm{out} & = ef\textrm{ e}^{-\Gamma_1\tau_1} \tag{\textit{N}=1}\\
I_\mathrm{out} & = ef\left(2\mathrm{e}^{-\Gamma_2\tau_2}+\frac{\Gamma_2}{\Gamma_1-\Gamma_2}
(\mathrm{e}^{-\Gamma_2\tau_2}-\mathrm{e}^{-\Gamma_1\tau_1})\right) \tag{\textit{N}=2}\\
I_\mathrm{out} & =
ef\left[3\mathrm{e}^{-\Gamma_3\tau_3}+\frac{2\Gamma_3}{\Gamma_2-\Gamma_3}(\mathrm{e}^{-\Gamma_3\tau_3}-\mathrm{e}^{-\Gamma_2\tau_2})\right. \nonumber\\
&{}+\frac{\Gamma_2\Gamma_3}{\Gamma_2-\Gamma_3}
\left(\frac{1}{\Gamma_1-\Gamma_3}\mathrm{e}^{-\Gamma_3\tau_3}-\frac{1}{\Gamma_1-\Gamma_2}\mathrm{e}^{-\Gamma_2\tau_2}\right. \nonumber\\
&\left.\left.{}+\frac{\Gamma_2-\Gamma_3}{(\Gamma_1-\Gamma_2)(\Gamma_1-\Gamma_3)}\mathrm{e}^{-\Gamma_1\tau_1}\right)\right] \tag{\textit{N}=3}\\
\end{align*}
The assumption of exactly $N$ initial electrons is not perfect, as imperfect quantisation in the SAW current leads to some dynamic QDs having $N+1$ or $N-1$ electrons \cite{Robinson2005}, and there is a small possibility that electrons may be transferred between adjacent dynamic QDs after initialisation. However, both of these processes should only affect a small percentage of dynamic QDs, and because $I_\mathrm{in}=Nef$ there must be equal numbers of $N+1$ and $N-1$ dynamic QDs whose effects would tend to cancel each other out, so the errors caused by this assumption should be less than the measurement errors in our system. Using these equations, the values of $\Gamma_n\tau_n$ are calculated as a function of barrier gate voltage in Fig.~\ref{Gtdp}. The tunnelling rate
is varied over an order of magnitude by a single gate, which shows great promise for making future SAW quantum devices.

The data in Fig.~\ref{Gtdp} are fitted using the analytical solution for the transmission probability of non-interacting electrons through the
saddle-point potential $V(x,y)=V_\mathrm{0}-\frac{1}{2}m^*\omega_x^2+\frac{1}{2}m^*\omega_y^2$ \cite{Buttiker1990, Connor1968}:
\begin{eqnarray*}
T_{i,j} & = & \delta_{i,j}\frac{1}{1+\mathrm{e}^{-\pi \epsilon}}\\ \textrm{where }\epsilon & = & \frac{2 \left[ E_n-\hbar
\omega_y\left(i+\frac{1}{2}\right)-V_\mathrm{0}\right]}{\hbar \omega_x}\\
\end{eqnarray*}
$V_\mathrm{0}$ is the potential at the centre of the barrier, $m^*$
is the effective mass of the electron, $\omega_x$ ($\omega_y$)
controls the curvature of the barrier perpendicular (parallel) to
the barrier, $\delta_{i,j}$ is the Kronecker delta function, $E_n$
is the energy of the incident electron and, assuming the electron
tunnels through the one-dimensional ground state, the sub-band index
$i=0$. The transmission probabilities are converted to tunnelling
probabilities by multiplying by a free parameter which describes the
number of attempts the electron makes at tunnelling in the time
$\tau$, and the other terms in the expression can be related to
changes in the tunnel barrier voltage ($V_\mathrm{T}$) by assuming a
simple capacitor model: $\Delta V_0=\alpha_{V_\mathrm{0}}\Delta
V_\mathrm{T}$ and $\frac{1}{2}m^*\Delta \omega_x^2 =
\alpha_{\omega_x} \Delta V_\mathrm{T}$ where each $\alpha$ is a
constant relating the coupling of the gate to the barrier potential;
$\omega_y$ is determined by the SAW potential amplitude and so
remains constant. We estimate $\alpha_{V_\mathrm{0}}=0.62\pm0.01$ by
applying a bias potential to the 2DEG until a breakdown current
starts to flow through the upper channel, which is expected to occur
when the Fermi energy of the 2DEG is level with the top of the
barrier. From the fitting parameters in table~\ref{tab}, we can
extract the addition energies $\Delta E_{n \rightarrow n+1}$ for an
$n$ electron dynamic QD. We find $\Delta E_{1 \rightarrow
2}=2.6\pm0.4\,\mathrm{meV}$ and $\Delta E_{2 \rightarrow 3}=14.1 \pm
1.3 \,\mathrm{meV}$ (these errors are from the fitting; there may be
other errors caused by the assumptions in the model that have not
been accounted for).
\begin{table}
\begin{tabular}{|c||c|c||c|}
\hline
&$\frac{2E_n-\hbar\omega_y}{\hbar\sqrt{2\alpha_{\omega_x}/m^*}}\,\mathrm{(V^{\frac{1}{2}})}$&$\frac{2\alpha_{V_\mathrm{0}}}{\hbar \sqrt{2\alpha_{\omega_x}/m^*}}\,\mathrm{(V^{-\frac{1}{2}})}$&$E_n-\frac{1}{2}\hbar\omega_y\,\mathrm{(meV)}$\\
\hline $n=1$&$0.0013\pm0.0004$&\multirow{3}{*}{$3.050\pm0.017$}&$0.27\pm0.08$\\
$n=2$&$0.015\pm0.002$&&$2.9\pm0.4$\\
$n=3$&$0.084\pm0.006$&&$17.0\pm1.2$\\ \hline
\end{tabular}
\caption{Fitting parameters from Fig.~\ref{Gtdp}, used to derive the addition energies of the dynamic QD.} \label{tab}
\end{table}

The energy of the dynamic QD will be increased by a Coulomb repulsion when adding an electron to the dot. The constant interaction model of a
QD predicts $\Delta E_{n \rightarrow n+1} = e^2/2C+\delta E_\mathrm{sp}$ with a capacitance $C$, at equal gate voltages and where $\delta E_\mathrm{sp}$ is the single-particle energy spacing (for a discussion of Coulomb energies within QDs, including the limitations of this constant-interaction model,
see Ref.~\cite{Kouwenhoven2001}). This predicts the ratio $\Delta E_{2 \rightarrow 3}/\Delta E_{1 \rightarrow 2} \approx 1$, whereas we find $\Delta
E_{2 \rightarrow 3}/\Delta E_{1 \rightarrow 2} = 5.4 \pm 1.0$. The difference is too large to be attributed solely to the single-particle energy---we suggest that the large variation in addition energies may be due to the complexities of the exchange and Coulomb interactions in few electron QDs which would require a self-consistent theory of electron-electron interactions to model accurately (note that the distance from QD to reservoir 2DEG is greater in our dynamic QDs than in previous static QD measurements, which will reduce the screening of the Coulomb interaction by the reservoir and could result in larger electron-electron effects).
However, the very large discrepancy may also suggest that assumptions in the saddle-point
tunnelling model (e.g.~ignoring electron-electron interactions in the tunnelling process or assuming the rate is only sensitive to the potential
at the tunnel barrier) are affecting the calculation, but while our measured addition energies may contain inaccuracies due to the approximations
incorporated into our model, we note that the energies are of comparable order of magnitude to those measured in static few-electron quantum dots \cite{Tarucha1996, Ciorga2000}.

In summary, we have demonstrated observations of tunnelling on a $\sim 600\,\mathrm{ps}$ timescale by confining electrons in dynamic QDs using a
SAW. Tunnel rates may be determined from the currents flowing through the device by using rate equations. The tunnel rates are dependent on the
barrier voltage applied and on the number of electrons in the dot; fitting these dependencies to a saddle point tunnelling model gives addition
energies which we attribute to the Coulomb interaction. The physical behaviour of electrons confined to dynamic QDs is found to be similar to that of electrons in static QDs, indicating that dynamic QDs can provide an additional method of probing the fundamental behaviour of electrons in QDs.

This work was part of the QIP IRC www.qipirc.org (Grant~No.~GR/S82176/01). MRA thanks Toshiba Research Europe Ltd.\ and UK EPSRC for funding.
Calculations of device potentials were carried out using the GatesCalc program written by A.~L.~Thorn.

\end{document}